\documentclass[10pt,conference]{IEEEtran} 
\IEEEoverridecommandlockouts
\usepackage{cite}
\usepackage{amsmath,amssymb,amsfonts}
\usepackage{algorithmic}
\usepackage{graphicx}
\usepackage{textcomp}
\usepackage{xcolor}
\usepackage{multirow}
\def\BibTeX{{\rm B\kern-.05em{\sc i\kern-.025em b}\kern-.08em
    T\kern-.1667em\lower.7ex\hbox{E}\kern-.125emX}}
\begin{document}

\title{Distribution Awareness for AI System Testing}

\author{\IEEEauthorblockN{David Berend}
\IEEEauthorblockA{
\textit{Nanyang Technological University, Singapore}\\
bere0003@e.ntu.edu.sg}
}

\maketitle

\begin{abstract}
As Deep Learning (DL) is continuously adopted in many safety critical applications, its quality and reliability start to raise concerns. Similar to the traditional software development process, testing the DL software to uncover its defects at an early stage is an effective way to reduce risks after deployment. Although recent progress has been made in designing novel testing techniques for DL software, the distribution of generated test data is not taken into consideration. It is therefore hard to judge whether the identified errors are indeed meaningful errors to the DL application. Therefore, we propose a new distribution aware testing technique which aims to generate new unseen test cases relevant to the underlying DL system task. Our results show that this technique is able to filter up to 55.44\% of error test case on CIFAR-10 and is 10.05\% more effective in enhancing robustness.
\end{abstract}


\section{Introduction}


Recently, deep learning (DL) achieved tremendous success with an increasing demand for automation and intelligent support for safety-critical areas, such as autonomous driving~\cite{auto_shanghai, auto_california} and healthcare~\cite{medi_cancer}, where quality assurance is of special importance. We have already witnessed the accidents and negative social impacts that were caused by quality issues of DL software, e.g., Tesla/Uber accidents~\cite{acc1, acc2}, wrong diagnosis in healthcare, e.g. cancer or diabetes~\cite{medi_cancer}. Therefore, systematic testing to uncover the incorrect behavior and understand the capability of the DL software is pressing and important.

Deep-learning follows a data-driven programming paradigm which is different from traditional software whose decision logic is mostly programmed by the developer. Therefore, new ways for software testing have been proposed for the DL domain~\cite{test:deepxplore,test:deepgauge,test:tensorfuzz,test:surprise}. Here, test cases are generated by applying mutations to the data under test~\cite{test:deepxplore, test:deeptest, test:concolic, test:deeproad, test:deephunter, test:tensorfuzz, test:deepstellar}, which however still lacks interpretation on the \emph{detected errors}. 

The fundamental assumption of deep learning is that the training data follows some \textit{distribution}, the In-Distribution (ID), which is aligned with the task the DL system tries to solve. When a testing framework produces a new error test cases, it remains currently unclear if such error is caused by a defect of the DL system or if the error test case follows a different distribution and is not relevant to the DL task, defined here as out-of-distribution (OOD). Thereby, the root cause of an error may be identified through analyzing its distribution behavior, which makes us rethink how to apply state-of-the-art testing frameworks. The challenge of OOD detection is that there is no perfect ground truth. Thereby, related work tries to utilize model behavior analysis to understand when an input is in fact OOD for which promising results have been achieved~\cite{ood:baseline, ood:oe, ood:odin, ood:maha}. 

Current testing frameworks use coverage criteria to guide the test case generation in identifying new unseen data. To bridge the gap from data distribution to DL testing activities, we present the first distribution-guided coverage criteria to guide the test case generation to new unseen data while providing a higher guarantee to the validity of the identified errors to DL system task.
More specifically, the new OOD-guided testing technique is evaluated on three different testing frameworks~\cite{test:deeptest, test:deepxplore, test:deephunter} with two different OOD-integrated coverage guided criteria and on two commonly used benchmark datasets CIFAR-10 and MNIST~\cite{ood:baseline, ood:oe}.

Our novel OOD-guided coverage criteria is able to detect up to 55.44\% of errors on the CIFAR-10 dataset which are not relevant to domain. When retraining the DNN model with the distribution-relevant errors, the DNN model is 10.05\% more accurate than the DNN model retrained with errors unaware of the distribution. Even more critical is the observation that when the DNN model is retrained with OOD errors only, the accuracy decreases on average by 54.62\%. The results reflect the importance of distribution awareness and calls for attention when designing future DL testing techniques.

\section{Background \& Related Work} 
\textit{DL testing.} Currently, quite a few techniques~\cite{test:deepxplore, test:deeptest, test:deeproad, test:deephunter, test:deepstellar, test:deepgauge, test:concolic, test:tensorfuzz} are proposed to test the new data-driven DL software. Coverage-guided testing is a representative and widely used technique, which usually contains three main components: the \textit{data mutation operator} to generate diverse test cases, the \textit{coverage criteria} to measure the degree of how much the DNN is tested, and the \textit{oracle} to judge whether a new test case is a benign test case, {(i.e., correctly predicted),} or an error test case, {(i.e., incorrectly predicted)}.

\textit{OOD detection.} Given two datasets $A$ and $B$, which follow the data distribution of $D_A$ and $D_B$, respectively, a DNN is trained on $A$.
If $A$ and $B$ have similar distributions, the well trained DNN is more likely to handle data from $B$ correctly. If they have a totally different distribution (e.g., cars and airplane images), the DNN is not expected to handle the data from $B$. OOD methods calculate an OOD score for a new input. If the score is below the defined threshold, it is ID. Otherwise, it is OOD.  
Some OOD detection has been recently proposed to address the high-dimensional issues, such as \cite{ood:llratio, ood:llratio2, ood:odin, ood:gen_ensemble, ood:scale_ensemble, ood:reject_class, ood:conf_cal, ood:conf_cal, ood:baseline, ood:cos_sim, ood:typicality, ood:maha, ood:oe}. These techniques provide different ways to evaluate the distribution of training data. This work inherits those techniques and identifies the best suited one, Outlier Exposure~\cite{ood:oe}, to integrate into existing coverage criteria methodologies.
\begin{figure}[t!]
    \includegraphics[page=1, width=\columnwidth]{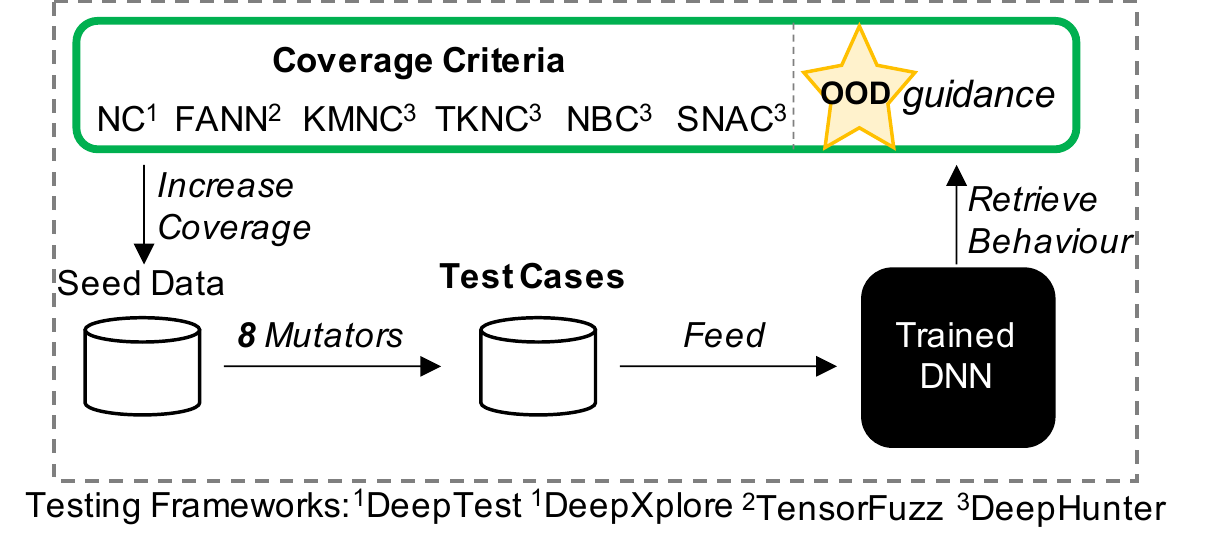}
    \vspace{-20pt}
    \caption{Testing workflow with OOD guidance}
    \label{fig:overview}
    \vspace{-4mm}
\end{figure}

\section{Methodology}\label{sec:method}

Fig.~\ref{fig:overview} shows the overview of the AI system testing procedure together with the novel OOD guided coverage criteria proposed by this work. Before OOD is integrated into the coverage criteria the optimal method is identified following commonly used evaluation procedures presented by related work in which OOD benchmark datasets are utilized~\cite{ood:baseline, ood:oe, data:omniglot, data:fashionmnist}. Afterward, the OOD method is integrated into the testing cycle. Here, seed data is taken from the training set on which the AI system is trained. Then, eight common data mutation operators including affine and color transformations are applied to the seeds individually, which produces test cases. These test cases are given to the DNN model as input. Afterward, the prediction outcome is analysed together with the coverage analysis joined by the OOD-guided neural behavior analysis. If the test case is predicted correctly and increases coverage while following the trained distribution (unseen in-distribution data) it is placed back into the seed data pool. If the test case resulted in an error, the error is analysed using the OOD-coverage criteria to see if it is relevant to the underlying system task. After going through the testing cycle for a defined amount of iterations, we compare the identified benign and error test cases with and without the OOD-guidance for the presented coverage criteria. 
Finally, we use the error test cases for retraining the DNN model to assess their effectiveness in predicting a separate test set consisting of errors which all frameworks identified and were not used for retraining. 

In this paper we present the results of the coverage criteria Neural Coverage (NC) used by DeepXplore and DeepTest~\cite{test:deepxplore, test:deeptest} and k-Multisection Neuron Coverage (KMNC) from DeepGauge\cite{test:deepgauge} used by DeepHunter~\cite{test:deephunter}. NC measures the total activation of neurons compared to the training data while KMNC applies a more fine grained analysis measuring the activation k-activation ranges. To integrate the OOD-methodology into the coverage criteria we retrieve the OOD-score following the state-of-the-art OOD methods and retrieve the OOD-score distribution from the training data. If a new test case has an OOD-score exceeding a defined threshold we define it as OOD. We choose the 99-percentile of the In-Distribution as OOD-threshold. To test the effectiveness of the OOD-guided testing we retrieve 2000 error test cases which we use as testset. Afterward, we retrain the DNN model on the original training set including 10.000 seperate identified error test cases. We investigate three settings: Errors selected at random as baseline, errors which follow the trained distribution (our enhancement) and only OOD-errors as additional comparison.

\section{Evaluation}
For our evaluation we choose the two benchmark datasets MNIST and CIFAR-10 which are trained on commonly used DNN model architectures LeNet-5~\cite{dnn:lenet5} and DenseNet-121~\cite{dnn:densenet}, respectively, together with best performing OOD-methodology Outlier-Exposure~\cite{ood:oe} for coverage criteria integration. 
For each DNN model we evaluate how many benign and error test cases are retrieved before and after our coverage criteria enhancement. The column \textit{before} in Table \ref{tab1} represents the traditional coverage criteria and \textit{after} represents the OOD-guided coverage criteria enhancement. Finally, column \textit{In \%} shows the ratio of test cases which are declared as OOD and thereby likely to be irrelevant to the DL system task. The results show that for NC, more OOD test cases have been identified for KMNC which makes sense, as KMNC is more fine grained in assessing the neural activation when selecting test cases. However, even KMNC shows a difference of up to 9.76\% of error test cases which the OOD-guided enhancement was able to filter. Finally, to assess the effectiveness of the OOD-guided enhancement we retrain the DNN model as described in Section \ref{sec:method}. Here, we identify that when retrained with errors selected by our novel technique the DNN model is 10.05\% more accurate on the testset than the DNN model retrained on errors unaware of the distribution. Furthermore, when retrained with OOD errors only, the accuracy decreases on average by 54.62\% further showcasing the importance of distribution awareness for DL system testing. 

\section{Discussion \& Future Work}
DL  testing  tools  should  be  aware  of distribution. A promising direction is to develop more fine-grained distribution-aware criteria which further incorporate the OOD-methodology into their own calculation of coverage. A future research direction is to further analyze the root cause of ID and OOD errors,  which can provide guidance for \textit{repairing} the model from a data and DNN architecture perspective under regard of the presented threshold of this work.

\begin{table}[t!]
\caption{OOD-guided coverage criteria results}
\centering
\resizebox{1.0\columnwidth}{!}{
\begin{tabular}{cccccccc}
\hline \hline 
\multirow{2}{*}{Dataset} & \multirow{2}{*}{Type} & \multicolumn{3}{c}{NC} & \multicolumn{3}{c}{KMNC}\tabularnewline
 &  & Before & After & $\Delta\%$ & Before & After &  $\Delta\%$\tabularnewline
\hline \hline 
\multirow{2}{*}{CIFAR-10} & Benign & 1113 & 425 & 61.81\% & 59255 & 58932 & 0.55\%\tabularnewline
 & Error & 24646 & 10983 & 55.44\% & 3656 & 3299 & 9.76\%\tabularnewline
 \hline 
\multirow{2}{*}{MNIST} & Benign & 999 & 208 & 79.18\% & 16370 & 16413 & -0.26\%\tabularnewline
 & Error & 12332 & 10757 & 12.77\% & 2395 & 2210 & 7.72\%\tabularnewline
\hline \hline 
\end{tabular}
}
\label{tab1}
\vspace{-10pt}
\end{table}

\bibliographystyle{IEEEtran}
\bibliography{IEEEabrv,lib}
\vspace{12pt}

\end{document}